\documentstyle[epsf,multicol,aps]{revtex}
\begin{document}

\title{
Interaction-induced localization-delocalization transition
in the double-layer quantum Hall system
}\par

\author{
Jun-ichiro Watanabe$^{\dagger, \ast}$
and Tatsuya Nakajima$^{\dagger \dagger}$
}\par

\address{
Physics Department, Graduate School of Science, Tohoku University,
Sendai 980-8578, Japan 
}\par

\date{\today}

\maketitle

\begin{abstract}
We report on numerical studies of the energy spectrum
and the localization properties
in the double-layer quantum Hall system
at $\nu = 1$. 
The Coulomb interaction is treated
by the Hartree-Fock approximation,
and the localization properties
in the presence of disorder are studied
by evaluating participation ratios for
the Hartree-Fock eigenfunctions.
We show that
the extended states seem to exist
only near each center of the two subbands split
by the exchange-enhanced energy gap.
It is also shown that
the self-consistent orbitals whose energies are 
close to the Fermi energy
appear to become extended together with
the reduction in the energy gap
as the layer separation increases.
The collapse of the energy gap expected from our results
is consistent with
the incompressible-compressible transition observed in
recent experiments,
and the change of the localization properties
near the Fermi energy can explain the disappearance
of the quantum Hall effect for large layer separations 
very well.

\end{abstract}



\begin{multicols}{2}
\narrowtext

\section{Introduction}

When the integer quantum Hall effect (QHE) is studied theoretically,
it is usually assumed without justification that
the Coulomb interaction between electrons can be safely ignored.
In a strong magnetic field,
the eigenfunctions of the single-particle Hamiltonian
for two-dimensional non-interacting electron systems
are localized by a disorder potential at almost all energies
except for a discrete set of critical 
energies $\{ \varepsilon _{c N} \}$
near the center of each disorder-broadened Landau level.
Theoretical studies suggest that at $T = 0$
the Hall conductivity jumps by $e^2/ h$
each time the Fermi energy $\varepsilon _F$ crosses
one of the critical energies,
and that the longitudinal conductivity is
zero if $\varepsilon _F \neq \varepsilon  _{c N}$.
These suggestions are supported
by many experimental studies \cite{Huck}.

However, such localization properties are not always guaranteed
when two different Landau levels are nearly degenerate.
In fact, the numerical studies of 
the double-layer quantum Hall (QH) system
in the absence of electron-electron interactions \cite{Ohtsuki,Sorensen} 
could not obtain the reasonable localization properties 
in case of nearly degenerate Landau levels
({\it i.e.}, in the weak interlayer-tunneling case).
This is a typical case where the Coulomb interaction should 
be considered even for understanding 
the integer QHE qualitatively.
Thus we consider this system
in the presence of interactions in this paper.
We show that 
the exchange-enhanced energy gap \cite{fogler} 
appears in this interacting system 
and that 
the localization properties consistent with
the observation of the QHE can be obtained 
for small layer separations 
even in the weak interlayer-tunneling case.

In double-layer QH systems \cite{review},
the interlayer-tunneling of electrons
brings about the mixing of the Landau levels 
in the two layers, and 
the Landau levels split 
into symmetric and antisymmetric combinations
about the center of the double-layer structure.
The energy gap, $\Delta _{\rm SAS}$, between them
is proportional to the tunneling amplitude,
and it needs to be small for the nearby degeneracy 
of the Landau levels.

Such samples are realized experimentally
and the transport 
properties have been investigated \cite{Boebinger,Murphy}.
In these experiments using high-mobility samples,
it has been reported that the Coulomb interaction plays
an important role on the ground-state properties and low-lying
excitations.
For $\nu = 1$, 
the phase diagram against the layer separation, $d$,
which controls the strength of 
the interlayer interactions,
and $\Delta _{\rm SAS}$ is obtained
experimentally \cite{Murphy}.
The phase diagram shows 
that the QH state disappears for $d > d_c$ and that
the critical separation $d_c$ increases
as $\Delta _{\rm SAS}$ increases.

In a more recent experiment using high mobility samples with
weak interlayer-tunneling,
the zero in the longitudinal resistivity is replaced
for $d > d_c$ by a broad minimum similar to that observed
in the single-layer QH system at $\nu = 1/2$ \cite{hamilton}.
This suggests a transition from an incompressible QH state
with strong interlayer correlations
to compressible state consisting of two (weakly correlated) layers, 
where the metallic states of composite fermions are formed.

Theoretically the pseudospin formalism is often introduced
to describe the layer degrees of freedom in double-layer systems.
This is done by assigning the upper/lower layers
to the pseudospin $\uparrow/\downarrow$.
At $\nu = 1/m$ \,($m$: odd integer), the pseudospin ferromagnetism 
results from the interlayer-tunneling and 
exchange interactions between electrons \cite{review}.
The phase boundary between the QHE and the non-QHE phase 
was determined theoretically by assuming that the QHE phase
is destroyed together with the collapse 
of the pseudospin ferromagnetism \cite{Platz,Nakajima}.
The pseudospin-ferromagnetic ground state is shown
to evolve continuously from
tunneling-dominated to correlation-dominated
as $\Delta _{\rm SAS}$ decreases \cite{sss}.

The Hartree-Fock calculations have also been done
to study the QH systems, and
this approximation is expected to
describe the electronic properties well
especially for the integer filling factors.
In fact, this approximation has been used for the study of the 
double-layer $\nu = 1$ QH system in the absence 
of random disorder potential \cite{Cote1,moon1}.

In this paper,
we investigate the energy spectrum and 
the localization properties
in disordered double-layer QH systems at $\nu = 1$.
The Coulomb interaction is treated
by the Hartree-Fock approximation,
and the localization properties are studied
by evaluating participation ratios for
the Hartree-Fock eigenfunctions.
This method has been used in the studies 
on the interaction effects in
the single-layer QH system \cite{Eric,Eric2}.
We show that
the localization properties change qualitatively
because of the interaction effects \cite{cap}.

Our paper is organized as follows.
In Section II, we explain the model and calculation methods
which we use in this study.
In Section III, we first discuss the results of 
the numerical calculations in the absence of Coulomb interactions.
Through this discussion about 
the previous results \cite{Ohtsuki,Sorensen} and ours, 
it is shown that the Coulomb interaction should 
be considered to understand
the $\nu = 1$ QHE in double-layer system
with weak interlayer-tunneling.
After this discussion, we show our numerical results 
in the presence of Coulomb interactions.
These are our main results in this paper,
and it is shown that
the localization properties change together with
the reduction in the exchange-enhanced energy gap
as the layer separation increases.
Finally in Section IV, we briefly summarize our findings.

\section{Model and Method}

We consider a double-layer system of spin-polarized electrons in
a strong magnetic field perpendicular to the layers.
In double-layer systems,
there exists the interlayer-tunneling of electrons.
The single-particle wavefunctions then split into symmetric and
antisymmetric ones about the center of the double-layer structure,
and the energy gap between them, $\Delta _{\rm SAS}$, enters
as an energy scale.
The thickness of the wavefunction in each layer
is neglected for simplicity.

The two-dimensional coordinates in the two parallel planes
are denoted by ${\bf r} = (x, y)$,
and the layer degrees of freedom are described by
the pseudospin $\sigma = \ \uparrow$, $\downarrow$.
The Coulomb interaction between electrons
is then dependent on pseudospin $\sigma$
for a finite layer separation $d$.
Its Fourier transform $V_{\sigma \sigma ^{\prime}} (q)$ is
$2 \pi e^2/ \epsilon q$ if $\sigma = \sigma ^{\prime}$
({\it i.e.}, for the intra-layer interaction) and
$(2 \pi e^2/ \epsilon q) \,e^{- q d}$ 
if $\sigma \neq \sigma ^{\prime}$
({\it i.e.}, for the inter-layer one), where $\epsilon$ is 
the dielectric constant
of the host material.
The Coulomb interaction is treated self-consistently
within the Hartree-Fock (HF) approximation.

In the strong-magnetic-field limit,
it is enough to consider only the lowest Landau level
because one can neglect the Landau level mixing
by interactions or disorders.
The real spin degrees of freedom is also ignored
by assuming 
the spin polarization due to the Zeeman energy.
Our attention is restricted to this strong-field limit.
We apply the periodic boundary condition
to the single-particle wavefunctions
inside the two parallel rectangles
of dimensions $L_x$, $L_y$,
and use the Landau gauge ${\bf A} ({\bf r}) = (0, B x, 0)$.
One can then use the following set of basis functions
for the lowest Landau level:
\begin{eqnarray}
\phi_{j}({\bf r}) &=&
\bigg ( \frac{1}{L _y \sqrt {\pi} \,\ell} \bigg ) ^{1/2}
\nonumber \\
& & \times 
\sum _{k = - \infty}^{\infty}
\exp \biggl [ \,
i \,\frac{X_j + k L_x}{\ell ^2} \,y
- \frac {(X_j + k L_x - x)^{2}}{2 \ell ^{2}} \,\biggl ] ,
\label{eqn:LWF}
\end{eqnarray}
where $\ell \equiv \sqrt {c \hbar / e B}$ is the magnetic length,
and $X_j = 2 \pi \ell ^2 j/ L_y$ is the center coordinate
of the $j$-th Landau orbit \cite{DYoshioka}.
The orbital degrees of freedom in each layer 
are described by this set of the Landau orbits,
and the two layers can be distinguished
by the pseudospin index $\sigma$. 
Thus the set $\{ | j \sigma \rangle \}$
($j = 1, 2, \cdots ,N_{L}$, $\sigma = \ \uparrow , \downarrow$)
can be used as a basis set for representing 
the Hartree-Fock Hamiltonian,
where $N_{L} = L_x L_y / 2\pi \ell ^2$ is
the Landau level degeneracy in each layer.

In terms of the set of basis functions, $\{ | j \sigma \rangle \}$,
the matrix element of the Hartree-Fock Hamiltonian
is given by
\begin{eqnarray}
\langle j \sigma  | H_{\rm HF} |  j^{\prime} \sigma ^{\prime} \rangle
&=&
- \frac {\Delta _{\rm SAS}}{2} \,
\delta _{j j^{\prime}}\,
\delta _{\sigma , - \sigma ^{\prime}}
+ \delta _{\sigma \sigma ^{\prime}} \,
\langle j \sigma  | v_{\rm imp} |  j^{\prime} \sigma \rangle
\nonumber \\
& & \quad \quad \quad \quad \quad
+ \langle j \sigma  | V_{\rm HF} |  j^{\prime} \sigma ^{\prime} \rangle ,
\label {eq: hfham}
\end{eqnarray}
where $\delta _{j j^{\prime}}$ is a usual Kronecker delta,
the first term in the right hand side of eq.(\ref {eq: hfham})
is due to the interlayer-tunneling.
The second term and the third one in eq.(\ref {eq: hfham})
result from the impurity scattering and the Coulomb interaction 
between electrons, respectively.
The amplitudes of interlayer impurity scatterings are neglected
because of their small values and for simplicity.

The Hartree-Fock single-particle equation is given by
\begin{equation}
H_{\rm HF} \,| \varphi _{\alpha} \rangle =
\varepsilon _{\alpha} \,| \varphi _{\alpha} \rangle ,
\label {eq: HFeq}
\end{equation}
where $\varepsilon _{\alpha}$ and 
$| \varphi _{\alpha} \rangle$ 
are an eigenvalue and corresponding 
eigenstate of this equation, respectively.
Because the third term, 
$\langle j \sigma  | V_{\rm HF} |  
j^{\prime} \sigma ^{\prime} \rangle$, 
in eq.(\ref {eq: hfham})
is dependent on the set, 
$\{ | \varphi _{\alpha} \rangle \}$, 
as seen in eq.(\ref {eq: hfcmat}),
this single-particle equation must be solved self-consistently.
This is done by diagonalizing the $2 N_L \times 2 N_L$ matrix,
$\langle j \sigma  | H_{\rm HF} |  
j^{\prime} \sigma ^{\prime} \rangle$,
in eq.(\ref {eq: hfham}) numerically
and solving eq.(\ref {eq: HFeq}) iteratively
until the self-consistency is achieved.
Among the obtained eigenstates, $\{ | \varphi _{\alpha} \rangle \}$,
in eq.(\ref {eq: HFeq}),
the $N$ lowest-energy ones are occupied in $N$-electron systems,
and $N = N_{L}$ in case of $\nu = 1$.

The matrix element of the impurity scattering,
which is the second term in eq.(\ref {eq: hfham}),
is given by
\begin{eqnarray}
\langle j \sigma | v _{\rm imp} | j^{\prime} \sigma \rangle
&=& \frac {1}{L_x L_y} \sum _{{\bf q}}
v_{\sigma} ({\bf q}) \,
\delta \bigg ( j - j^{\prime}, \frac {q_y L_y}{ 2 \pi} \bigg ) \,
\nonumber \\
& & 
\times 
\exp \biggl [
- \frac {q^2 \ell ^2}{4} + i
\bigg (
q_x  X_{j}
- \frac {q_x q_y \ell ^2}{2}
\bigg )
\biggl ] ,
\end{eqnarray}
where $v_{\sigma} ({\bf q})$ is the Fourier transform
of the impurity potential, $v_{\sigma} ({\bf r})$,
in the layer $\sigma$,
$\delta (j, \,j^{\prime})$ is $1$ if $j = j^{\prime}$ (mod $N_L$)
and $0$ otherwise.
The sum over the wave vector ${\bf q}$ is over
$q_x =  (2 \pi/ L_x) \,n_x$,
$q_y =  (2 \pi/ L_y) \,n_y$ \,($n_x$, $n_y$: integers),
because the periodic boundary condition is used.

Our model disorder consists of randomly located
$\delta$-function scatterers 
with a random strength
uniformly distributed between $-V_{0}$ and $V_{0}$.
The disorder potential is then given by
\begin{equation}
v_{\sigma}({\bf r}) = \sum _{i}
V_{i} ^{\sigma} \,
\delta ({\bf r} - {\bf R}^{\sigma}_{i}) ,
\label{eq: mdo}
\end{equation}
where $V^{\sigma}_{i}$ and ${\bf R}^{\sigma}_{i}$ are
the strength and position of the $i$-th impurity
in the layer $\sigma$, respectively.
There exist ${N_{\rm imp}}$ impurities in each layer,
and we assume that the disorder potentials
in the two layers are {\it uncorrelated}, 
{\it i.e.}, there is no correlation
about $\{ {\bf R}^{\sigma}_{i} \}$ and $\{ V^{\sigma}_{i} \}$
between the two layers \cite{Sorensen}.
For this model disorder,
the energy scale which characterizes the Landau subband width is
given by $\Gamma = (V_0^2 N_{\rm imp}/ \ell ^2 L_x L_y)^{1/2}$
\cite{ando}.
We choose to work with
$N_{\rm imp}/ N_L = 2 \pi \ell ^2 N_{\rm imp}/ L_x L_y = 5$,
and keep $V_0/ \ell ^2$ constant
in order to use $\Gamma$ as the unit of energy.

The matrix element of the Coulomb interaction,
$\langle j \sigma  | V_{\rm HF} |  j^{\prime} \sigma ^{\prime} \rangle$,
which is the third term in eq.(\ref {eq: hfham}), is given by
\par
\vspace*{0.3cm}
\noindent 
$\langle j \sigma | V_{\rm HF} | j^{\prime} \sigma 
^{\prime} \rangle$ \par

\begin{eqnarray}
&=&
\sum _{\alpha} \sum _{j_1, j_2} 
\theta (\varepsilon _F  - \varepsilon _{\alpha} ) \nonumber \\
& & \times \bigg [
\delta _{\sigma \sigma ^{\prime}} 
\sum _{\sigma ^{\prime \prime}} 
\langle j j_1 | V_{\sigma \sigma ^{\prime \prime}} | 
j^{\prime} j_2 \rangle \,
\langle \varphi _{\alpha} | j_1 \sigma ^{\prime \prime} \rangle \,
\langle  j_2  \sigma ^{\prime \prime} | \varphi _{\alpha} \rangle 
\nonumber \\
& & \quad \quad \quad \quad 
- \langle j j_1 | V_{\sigma \sigma ^{\prime}} | 
j_2 j^{\prime} \rangle \,
\langle \varphi _{\alpha} | j_1 \sigma ^{\prime} \rangle \,
\langle  j_2  \sigma | \varphi _{\alpha} \rangle 
\bigg ]
\nonumber \\
&=&
\sum _{{\bf q} \in {\rm B.Z.}}
e^{i q_x X_{j}} \,
\delta \bigg (j - j^{\prime}, \,\frac {q_y L_y}{2 \pi} \bigg ) 
\nonumber \\
& & \times \biggl [
\delta _{\sigma \sigma ^{\prime}} \sum _{\sigma ^{\prime \prime}}
\Delta _{\sigma ^{\prime \prime} \sigma ^{\prime \prime}} ({\bf q}) \,
U_{\rm H}^{\sigma \sigma ^{\prime \prime}} ({\bf q})
- \Delta _{\sigma ^{\prime} \sigma} ({\bf q}) \,
U_{\rm F}^{\sigma \sigma ^{\prime}} ({\bf q})
\biggl ] \label {eq: hfcmat} , 
\end{eqnarray}

\vspace*{-0.4cm}

\begin{eqnarray}
U_{\rm H}^{\sigma \sigma ^{\prime}} ({\bf q})
&=&
\frac {1}{2 \pi \ell ^2} \sum _{{\bf q}^{\prime} \neq {\bf 0}}
\delta \bigg ( \frac {q_x L_x}{2 \pi}, \,
\frac {q^{\prime}_x L_x}{2 \pi} \bigg ) \,
\delta \bigg ( \frac {q_y L_y}{2 \pi}, \,
\frac {q^{\prime}_y L_y}{2 \pi} \bigg ) \,
\nonumber \\
& & \quad \quad \quad \quad \quad \quad \quad \times 
V_{\sigma \sigma ^{\prime}} ( q^{\prime} ) \,
e^{- q^{\prime \,2} \ell ^2/ 2} , \\
U_{\rm F}^{\sigma \sigma ^{\prime}} ({\bf q})
&=&
\frac {1}{L_x L_y} \sum _{{\bf q}^{\prime} \neq {\bf 0}}
V_{\sigma \sigma ^{\prime}} ( q^{\prime} )
\nonumber \\
& & \times \exp \biggl [
- \frac { q^{\prime \,2} \ell ^2}{2}
+ i (q^{\prime}_x q_y - q^{\prime}_y q_x ) \ell ^2
\biggl ] , \\
\Delta _{\sigma \sigma ^{\prime}} ({\bf q})
&\equiv&
\frac {1}{N_L} \exp \left (
\frac{q^2 \ell ^2}{4}
- i \frac {q_x q_y \ell ^2}{2}
\right )
\langle \hat{\rho}_{\sigma \sigma ^{\prime}} ({\bf q}) \rangle
\nonumber \\
&=&
\frac {1}{N_L}
\sum _{j, j^{\prime}}
e^{- i q_x X_{j^{\prime}}} \,
\delta \bigg ( j^{\prime} - j, \,
\frac {q_y L_y}{2 \pi} \bigg ) \,
\nonumber \\
& & \times 
\sum _{\alpha} \theta (\varepsilon _F  - \varepsilon _{\alpha} ) \,
\langle \varphi _{\alpha} | j \sigma \rangle \,
\langle  j^{\prime}  \sigma ^{\prime} | \varphi _{\alpha} \rangle ,
\label {eq: alphasum} 
\end{eqnarray}

\noindent 
$\langle j_1 j_2 | V_{\sigma \sigma ^{\prime}} | 
j_3 j_4 \rangle $ \par
\begin{eqnarray}
&=&
\frac {\delta ( j_1 + j_2, j_3 + j_4 )}{L_x L_y} 
\sum _{{\bf q} \neq {\bf 0}}
V_{\sigma \sigma ^{\prime}} ( q ) \,
\delta \bigg ( j_1 - j_3, \,
\frac {q_y L_y}{2 \pi} \bigg ) \nonumber \\
& & \quad \quad \quad \quad \times 
\exp \biggl [
- \frac { q^2 \ell ^2}{2}
+ i q_x (X_{j_1} - X_{j_4} ) 
\biggl ] , \label {eq: twobo}  \\
\hat{\rho} _{\sigma \sigma ^{\prime}}({\bf q})
&=&
\int _{0}^{L_x} dx 
\int _{0}^{L_y} dy \
e^{- i {\bf q} \cdot {\bf r}} \,
\Psi ^{\dag}_{\sigma} ({\bf r}) \Psi _{\sigma ^{\prime}} ({\bf r}) ,
\end{eqnarray}
where the sum over ${\bf q}$ in eq.(\ref {eq: hfcmat}) is
over `the Brillouin zone',
$q_x =  (2 \pi/ L_x) \,n_x$, $q_y =  (2 \pi/ L_y) \,n_y$ \
($n_x$, $n_y = 1, 2, \cdots, N_L$),
and the quantities $U_{\rm H}^{\sigma \sigma ^{\prime}} ({\bf q})$
and $U_{\rm F}^{\sigma \sigma ^{\prime}} ({\bf q})$ correspond
to the Hartree and Fock potential, respectively \cite{Aers}.
The quantity $\Delta _{\sigma \sigma ^{\prime}} ({\bf q})$ is
proportional to the expectation value
of the density
operator $\hat{\rho} _{\sigma \sigma ^{\prime}}({\bf q})$.
The two-body matrix element, 
$\langle j_1 j_2 | V_{\sigma \sigma ^{\prime}} | 
j_3 j_4 \rangle$, in eq.(\ref {eq: twobo}) is determined by the
Fourier transform, 
$V_{\sigma \sigma ^{\prime}} ( q )$, of the Coulomb interaction,
and a field operator, $\Psi_{\sigma}({\bf r})$, 
for pseudospin $\sigma$
is 
considered within the subspace of the lowest Landau level.
In eq.(\ref {eq: hfcmat}) and eq.(\ref {eq: alphasum}), 
$\theta (x)$ and $\varepsilon _F$ are
the Heaviside step function and the Fermi energy, respectively,
and only the $N$ lowest-energy eigenstates in eq.(\ref {eq: HFeq})
contribute to the sum over $\alpha$.
About the
quantities $U_{\rm F}^{\sigma \sigma ^{\prime}} ({\bf q})$,
$U_{\rm H}^{\sigma \sigma ^{\prime}} ({\bf q})$,
and $\Delta _{\sigma \sigma ^{\prime}} ({\bf q})$,
it is enough to consider them only within `the Brillouin zone'
because of their periodicity.

When the HF single-particle equation 
is solved,
the quantity $\Delta_{\sigma \sigma ^{\prime}}({\bf q})$ 
can be used 
to check the self-consistency of the calculated results.
We have judged the convergence of the calculated ones
by the following condition
\begin{equation}
\delta \equiv \frac {1}{4 N_L^2}
\sum_{\sigma , \sigma ^{\prime}}
\sum _{{\bf q} \in {\rm B.Z.}}
| \Delta_{\sigma \sigma ^{\prime}} ^{k+1} ({\bf q})
- \Delta _{\sigma \sigma ^{\prime}} ^{k} ({\bf q}) |
< 10^{-6} ,
\end{equation}
where $k$ represents each iteration step.
We note that our calculations have been done under the constraint
that the average numbers of electrons are the same
in the two layers,
{\it i.e.}, $\Delta_{\sigma \sigma} ({\bf 0}) = \nu / 2 = 1/2$
for $\sigma = \ \uparrow$, $\downarrow$ \cite{Cote1}.

In order to investigate the localization properties, 
we evaluate the participation ratios for the self-consistent
Hartree-Fock eigenstates \cite{Eric2}.
The participation ratio is given by
\begin{equation}
P _{\alpha} \equiv
\biggl [ \,
L_x L_y
\int _{0}^{L_x} dx
\int _{0}^{L_y} dy \
| \varphi _{\alpha} ({\bf r}) | ^4 \,
\biggl ] ^{-1}
\end{equation}
for a normalized
eigenstate $\varphi _{\alpha} ({\bf r}) = \langle \,{\bf r}\,|
\varphi _{\alpha} \rangle$, and 
$P _{\alpha} \sim \xi _{\alpha}^2/ L_x L_y$,
where $\xi _{\alpha}$ is the localization 
length of an eigenstate, $\varphi _{\alpha} ({\bf r})$.
As the eigenstate becomes more extended,
the participation ratio becomes larger. 
That is, the participation ratio shows 
how extended the eigenstate is.

For our numerical calculations,
$\Delta _{\rm SAS}$
and $e^2/ \epsilon \ell$ can be used as
the energy scales for the interlayer-tunneling 
and the Coulomb interaction, respectively.
As the unit of energy,
the strength, $\Gamma$, of the impurity potential
is used,
and we consider the following cases, 
$\Delta _{\rm SAS}/ \Gamma = 0.1$,
$(e^2/ \epsilon \ell)/ \Gamma  = 20$, $d/ \ell = 1.2, 1.5, 1.8$. 
For these values of the parameters,
the amplitude of the interlayer-tunneling is very small
and the Coulomb interaction is much stronger than the disorder potential
($e^2/ \epsilon \ell \sim 100$K for typical GaAs samples).
Thus we can compare the calculated results with
the experimental ones for high mobility samples with
weak interlayer-tunneling.

In the process of our self-consistent calculations,
we first solve eq.(\ref {eq: HFeq}) in the absence
of Coulomb interactions, {\it i.e.}, for
$ (e^2/ \epsilon \ell)/ \Gamma = 0$,
where no self-consistency is required.
By increasing the ratio, $(e^2/ \epsilon \ell)/ \Gamma$,
gradually to the required value
and using the latest results as input data
in a new calculation step \cite{Eric},
we have obtained the self-consistent solution of eq.(\ref {eq: HFeq}).

We also show the results in the absence of Coulomb interactions,
because the comparison between them and those in the presence 
of interactions clarifies 
the importance of Coulomb interactions
for the localization properties in the double-layer QH system
at $\nu = 1$.
The calculations in the absence of interactions are done
for $\Delta _{\rm SAS}/ \Gamma = 1$ and $0.1$,
{\it i.e.}, in both strong and weak tunneling cases.

\section{Numerical Result and Discussion}
\subsection{In the absence of Coulomb interactions}

We first discuss the double-layer QH system
in the absence of Coulomb interactions.
The localization properties in this non-interacting 
system were studied previously \cite{Ohtsuki,Sorensen}.
Although our numerical results in the non-interacting 
case are almost similar to previous ones \cite{Sorensen},
we show them in Figure 1
to make our discussion easy to understand.
The importance of Coulomb interactions 
in double-layer QH systems can be understood
by comparing these results with ones in the presence of interactions,
which will be given in the next subsection.

We calculated the density of states (DOS) and participation ratio
for several impurity configurations
in the non-interacting double-layer QH system with the Landau level
degeneracy, $N_L = 256$.
Because the square ($L_x = L_y = L$) systems are now considered,
the dimension $L$ is given
by $L = \ell \sqrt {2 \pi N_L} \simeq 40 \ell$
($\ell$: the magnetic length).
Figure 1 shows the result for one of these impurity configurations
in the cases of $\Delta _{\rm SAS}/ \Gamma = 1$ and $0.1$.
In the non-interacting case,
the parameter, $\Delta _{\rm SAS}/ \Gamma$,
is needed to characterize the disordered double-layer
QH system. 
The layer separation, which controls interlayer interactions,
does not need to be specified
in spite of its importance in the real experiments.

In the figures of the DOS in Figure 1,
the numbers of eigenstates 
within a finite width of energy, $\Delta E$, are plotted.
The width $\Delta E$ is several times as large as
the average energy-level spacing.
We note that
the energy $E$ in the figures is given in units of $\Gamma$
all through this paper
and that each of the vertical broken lines
indicates the highest energy eigenvalue
among those of the occupied eigenstates.

For $\Delta _{\rm SAS}/ \Gamma = 1$, which corresponds to
the strong interlayer-tunneling case,
there exist two subbands
which mainly consist of symmetric and antisymmetric combinations
of isolated layer states, respectively.
The localization properties within each subband
are similar to those in the single-layer QH systems
as pointed out previously \cite{Ohtsuki,Sorensen}.
That is,
the eigenfunctions are extended
only near the center of each disorder-broadened subband.

For $\Delta _{\rm SAS}/ \Gamma = 0.1$, however,
the symmetric and antisymmetric subbands are not well developed
in the DOS, and
the localization properties are quite different from
those for $\Delta _{\rm SAS}/ \Gamma = 1$.
In fact, the peak in the participation ratios
has a much broader width than that 
for $\Delta _{\rm SAS}/ \Gamma = 1$,
and the participation ratios near the Fermi energy 
take much larger values 
than those for $\Delta _{\rm SAS}/ \Gamma = 1$.

In the previous study \cite{Sorensen},
the finite-size scaling method was also used
for the non-interacting system with weak interlayer tunneling.
Then it was claimed 
that the extended states exist only at the two energies,
which are split by somewhat larger than $\Delta _{\rm SAS}$,
rather than across a band of finite width between the low- and
high-energy mobility edges in the thermodynamic limit.
In the energy interval between the two extended-state energies,
however,
their numerical values of the localization length in
the weak-tunneling case
are much larger than those in the strong-tunneling case.
Moreover it is not clear from the system-size dependence
of their results whether
the extended states do not exist across a band of finite width
in the thermodynamic limit, either.
Thus their claim seems to be controversial
as far as one judges from their numerical results.
Therefore we consider the effects of the Coulomb interaction
ignored in their study,
and show that 
the extended states seem to exist
only near each center of the two subbands split
by the interaction.


\subsection{In the presence of Coulomb interactions}

In the presence of Coulomb interactions,
it is necessary to specify the layer separation, $d/ \ell$,
which controls interlayer interactions,
as well as the interlayer-tunneling amplitude,
$\Delta _{\rm SAS}/ \Gamma$,
in order to characterize the disordered double-layer QH system.
In this subsection,
we consider the double-layer QH system
with $\Delta _{\rm SAS}/ \Gamma = 0.1$, {\it i.e.},
the weak interlayer-tunneling case.
The localization properties in this case remain unclear
as seen in the previous subsection.
About the layer separation,
we consider the three cases of $d/ \ell = 1.2, 1.5, 1.8$.
By using these values,
we can see that the localization properties 
change qualitatively as the layer separation increases.

We calculated the DOS and participation ratio
for several impurity configurations
in the double-layer QH system with the Landau level
degeneracy, $N_L = 80$.
The dimension $L$ is then given
by $L = \ell \sqrt {2 \pi N_L} \simeq 22 \ell$.
Figure 2 shows the result for one of these impurity configurations
in the cases of $\Delta _{\rm SAS}/ \Gamma = 0.1$,
$d/ \ell = 1.2, 1.5, 1.8$.
In the figures of the DOS,
the numbers of eigenstates within
a finite width of energy, $\Delta E$, are plotted,
and the energy $E$ in the figures is given in units of $\Gamma$.
Each of the vertical broken lines
in the figures of the participation ratio
indicates the highest energy eigenvalue
among those of the occupied eigenstates.

Let us first consider the case of $d/ \ell = 1.2$ 
in Figure 2.
We can see that the DOS has a large 
energy gap near the Fermi energy.
The participation ratios take small values (nearly zero) 
at the edges of each subband,
and take much larger values around the center of each subband.
These localization properties are almost similar to those 
for $\Delta _{\rm SAS}/ \Gamma = 1$ in Figure 1,
and are quite different from those 
for $\Delta _{\rm SAS}/ \Gamma = 0.1$ in Figure 1.
This difference results from the electron correlation effects
and the energy gap in Figure 2 is due to 
the exchange interactions between electrons.
In fact, for small layer separations, $d/ \ell$,
the ground state is a pseudospin-ferromagnetic one \cite{sss},
and the symmetric and antisymmetric combinations
of isolated layer states are separated from
each other by an exchange-enhanced energy gap.
Then the localization properties within each subband are 
expected to be similar to those in the single-layer QH systems
and be consistent with the observation of the integer QHE.

For larger layer separations ($d/ \ell = 1.5$, $1.8$),
however, 
the localization properties become complicated.
The participation ratios for 
the electronic eigenstates whose energies are
close to the Fermi energy take 
larger values than those in case of $d/ \ell = 1.2$,
{\it i.e.}, they are more extended
than the ones for $d/ \ell = 1.2$.
Moreover, the energy gap seen in the DOS
seems to decrease 
as the layer separation increases.
This decrease in the energy gap corresponds to 
the theoretical results from the viewpoint of the
pseudospin ferromagnetism \cite{Platz,Nakajima}
that
the pseudospin-ferromagnetic order are broken gradually
with the increase of layer separation.
If the energy gap continues to decrease and eventually 
collapses with the increase of layer separation,
it is consistent with the incompressible-compressible transition
reported in recent experiments \cite{hamilton}.

Although the size-scaling calculations are needed to
make the localization properties clear,
we claim the existence of 
the extended states only near the center of each subband
in case of $d/\ell = 1.2$ from the following facts.
One of them is that
the dependence of 
such localization properties on the layer separation
is consistent with the transition observed experimentally 
between the QHE and the non-QHE phase 
\cite{Murphy,hamilton}.
The other is that 
the localization properties 
in case of $d/\ell = 1.2$
are almost similar to those
in case of $\Delta _{\rm SAS}/ \Gamma = 1$
in Figure 1, 
although the origins of the energy gap are different.
We expect that our claim will be confirmed 
by the size-scaling calculations.

We also performed numerical calculations in the cases of
$\Delta_{\rm SAS}/ \Gamma = 1$, $d/ \ell = 1.5, 1.8, 2.0, 2.3$.
Although these results are not shown graphically in this paper,
we obtained the results which are similar to those
for $\Delta_{\rm SAS}/ \Gamma = 0.1$.
The energy gap then survives
for larger layer separations than that in case of
$\Delta_{\rm SAS}/ \Gamma = 0.1$.
This corresponds to the experimental\cite{Murphy} 
and theoretical results \cite{Platz,Nakajima}
that the critical separation $d_c$ increases
as $\Delta _{\rm SAS}$ increases.

Thus 
the exchange-enhanced energy gap
under the pseudospin-ferromagnetic order
appears between the two subbands, and 
the existence of the extended states
only near the center of each subband seems to be realized
for small layer separations by the electron correlation effects.
Unfortunately our Hartree-Fock calculations 
in this paper are limited
to those for small system sizes
and relatively small layer separations,
because it is difficult to obtain 
the numerical convergence
for larger system sizes or larger layer separations.
Therefore, for the confirmation of our claim about
the localization properties in weak interlayer-tunneling case,
further numerical studies are needed for larger systems.
Especially the size-scaling calculations should be done
to discuss the change of the localization properties
quantitatively.

\section{Summary}

We investigated the disordered double-layer QH system 
at $\nu=1$ numerically, 
and then the Coulomb interaction was taken 
into consideration within the Hartree-Fock approximation.
We examined the density of states and the participation ratios
for the self-consistent eigenstates of the Hartree-Fock Hamiltonian.
By considering the interaction effects in this disordered system,
it was found 
that the extended states seem to exist
only near each center of the two subbands split
by the exchange-enhanced energy gap
for small layer separations.
It was shown that
the eigenstates whose energies are 
close to the Fermi energy
appears to become extended together with
the decrease in the energy gap
as the layer separation increases.
The collapse of the energy gap,
which can be expected from our results,
is also consistent with 
the incompressible-compressible
transition reported in recent experiments.
Especially the change of the localization properties
near the Fermi energy can explain the disappearance
of the QHE phase for large layer separations very well.

\begin{acknowledgments}
We would like to thank Professor Komajiro Niizeki for some useful 
comments and valuable discussions.
We are also grateful to Nobuhisa Fujita for valuable discussions
about the Coulomb gap.
One of us (J.W.) would like to thank Eigo Yagi and Masaki Iwasawa
for some useful advice on the numerical calculations.
One of us (T.N.) was supported by JSPS 
(Japan Society for the Promotion of Science) 
Postdoctoral Fellowships for Research Abroad (1999).

\end{acknowledgments}


\end{multicols}
\widetext

\vspace*{3cm}

\begin{figure}
\epsfysize=12cm
\begin{center}
\leavevmode \epsffile{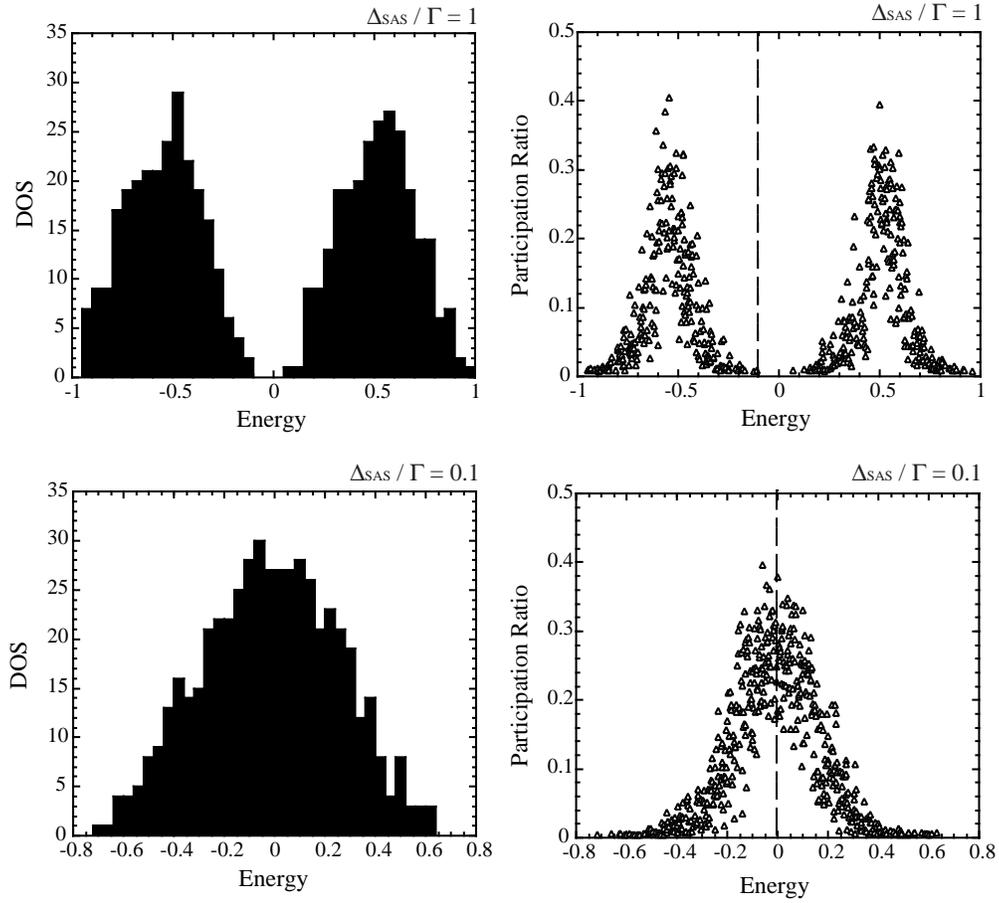}
\end{center}

\caption{The density of states (DOS) and participation ratio
in the non-interacting double-layer QH system at $\nu = 1$
are shown in both strong ($\Delta _{\rm SAS}/ \Gamma = 1$) 
and weak
($\Delta _{\rm SAS}/ \Gamma = 0.1$) interlayer-tunneling cases.
In the figures of the DOS,
the numbers of eigenstates within
a finite width of energy, $\Delta E$, are plotted.
Each of the vertical broken lines
in the figures of the participation ratio
indicates the highest energy eigenvalue
among those of the occupied eigenstates.
The Landau level degeneracy, $N_L$, is $256$,
and the energy $E$ in the figures is given
in units of $\Gamma$.}
\end{figure}

\begin{figure}
\epsfysize=18cm
\begin{center}
\leavevmode \epsffile{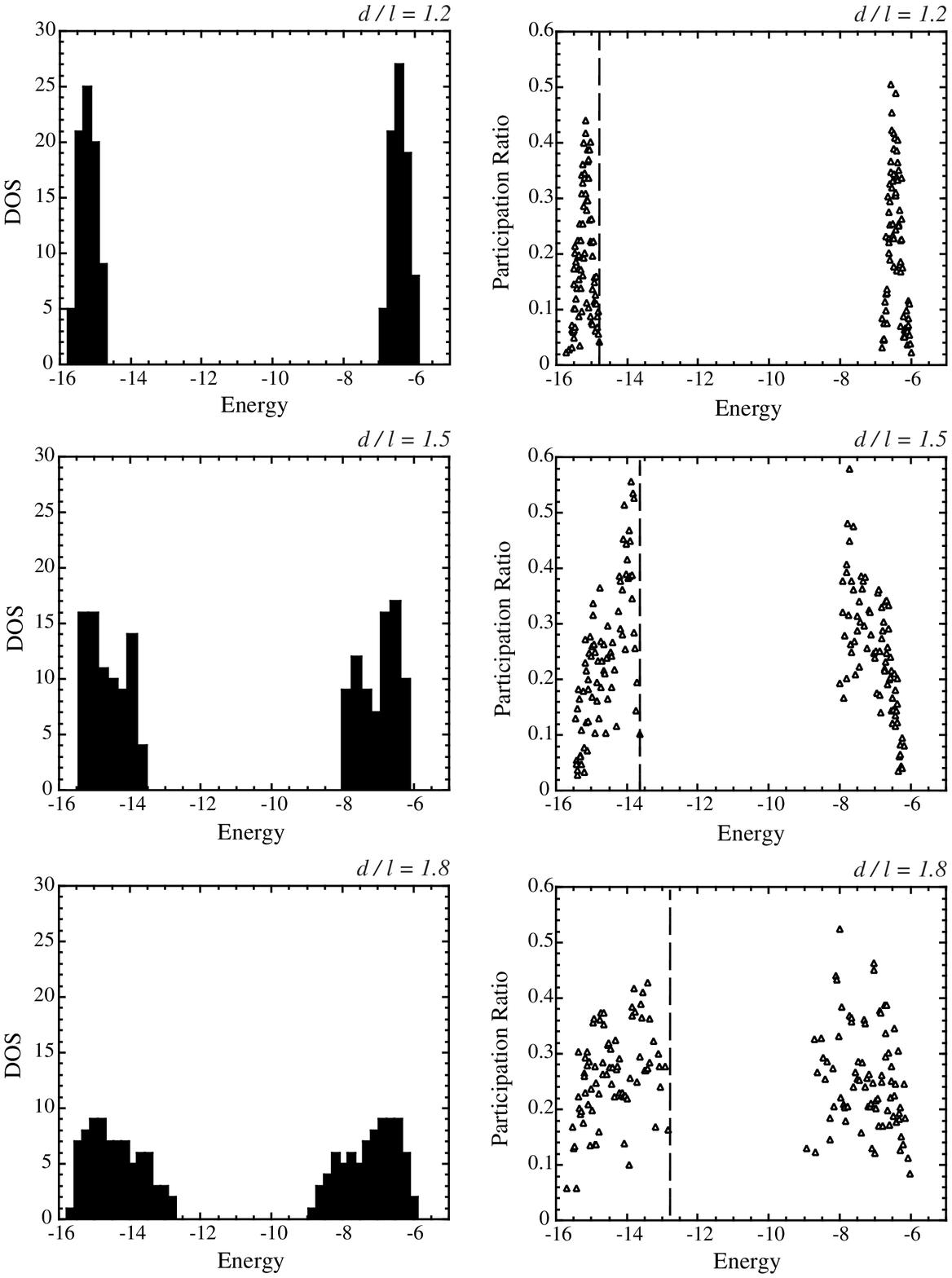}
\end{center}

\caption{The density of states (DOS) and participation ratio
for the self-consistent eigenstates
in the interacting double-layer QH system at $\nu = 1$ 
are shown in the weak interlayer-tunneling case
($\Delta _{\rm SAS}/ \Gamma = 0.1$).
As the layer separation increases,
the energy gap decreases and
the electronic eigenstates 
whose energies are close to the Fermi energy
seem to become extended.
The values used for the layer separation, $d/ \ell$, 
are $1.2$, $1.5$ and $1.8$,
and the Landau level degeneracy, $N_L$, is $80$.}
\end{figure}


\begin{thebibliography}{99}
\bibitem{cap}
In contrast to our study, 
the previous studies of the disordered QH system \cite{Eric,Eric2}
found no qualitative change of
the localization properties.
\end{thebibliography}
\end{document}